\documentclass[twocolumn,amsmath,amssymb,twoside,preprintnumbers,floatfix,showpacs,pre,superscriptaddress]{revtex4}

\usepackage{hyperref}
\usepackage[latin1]{inputenc}
\usepackage{vmargin}
\usepackage[english]{babel}
\usepackage{graphicx}
\usepackage{rotating}
\usepackage{wrapfig}
\usepackage[dvips]{epsfig}
\usepackage{amsmath}
\usepackage{amssymb}
\usepackage{bm}
\usepackage{dcolumn}
\usepackage{natbib}

\graphicspath{{}{images/}{submission/}}

\newcommand{\eq}[1]{Eq.\,(\ref{#1})}
\newcommand{\eqns}[2]{Eqns.\,(\ref{#1})--(\ref{#2})}

\newcommand{\fig}[1]{Fig.\,\ref{#1}}
 
\newcommand{\sect}[1]{Sec.\,\ref{#1}}
\newcommand{\be}{\begin{equation}}
\newcommand{\ee}{\end{equation}}
\newcommand{\bea}{\begin{eqnarray}}
\newcommand{\eea}{\end{eqnarray}}
\newcommand{\ba}[1]{\begin{array}{#1}}
\newcommand{\Ref}[1]{Ref.~\cite{#1}}
\newcommand{\Refs}[1]{Refs.~\cite{#1}}
\newcommand{\etal}{\emph{~et~al.~}}
\newcommand{\ea}{\end{array}}
\newcommand{\angst}{\,\mathrm{\AA}}
\renewcommand{\epsilon}{\varepsilon}
\renewcommand{\theta}{\vartheta}
\renewcommand{\vec}[1]{\mathbf{#1}}
\newcommand{\kvector}{\mbox{$\vec{k}$-vec}\-tor}
\newcommand{\lowkpeak}{\mbox{low-$\vec{k}$-peak}}
\newcommand{\highkpeak}{\mbox{$2^{nd}$-order-peak}}
\newcommand{\pH}{p\mathrm{H}}
\newcommand{\pHvalue}{\mbox{$\pH$-value}}
\newcommand{\Alumina}{Al$_2$O$_3$}
\newcommand{\mmoll}{\,\mathrm{mmol/l}}
\newcommand{\invsec}{/\mathrm{s}}

\begin{document}
\title{A Stability Diagram for Dense Suspensions of Model Colloidal \Alumina{}-Particles \mbox{in Shear Flow}}
\author{Martin Hecht}
\affiliation{Institute for Computational Physics,
Pfaffenwaldring 27,
70569 Stuttgart,
Germany}
\author{Jens Harting}
\affiliation{Institute for Computational Physics,
Pfaffenwaldring 27,
70569 Stuttgart,
Germany}

\author{Hans J. Herrmann}
\affiliation{Computational Physics, IFB, Schafmattstr. 6, ETH Z\"{u}rich,
CH-8093 Z\"{u}rich, Switzerland}
\affiliation{Departamento de F\'{i}sica, Universidade Federal do Cear\'{a}
Campus do Pici, 60451-970 Fortaleza CE, Brazil}

\date{\today}

\begin{abstract}
\noindent
\textbf{Abstract.}
In Al$_2$O$_3$ suspensions, depending on the experimental conditions very different 
microstructures can be found, comprising fluid like suspensions, a repulsive structure, 
and a clustered microstructure. For technical processing in ceramics, the knowledge
of the microstructure is of importance, since it essentially determines the stability 
of a workpiece to be produced. To enlighten this topic, we investigate these suspensions 
under shear by means of simulations. We observe cluster formation on two different 
length scales: the distance of nearest neighbors and
on the length scale of the system size. We find that the clustering behavior 
does not depend on the length scale of observation. If inter-particle interactions
are not attractive the particles form layers in the shear flow. The results are
summarized in a stability diagram.
\end{abstract}



\maketitle

% ==========================================================================================
\section{Introduction}
% ------------------------------------------------------------------------------------------
\noindent

\label{chap_Intro}
Colloid science is a very fascinating research field, gaining more and more
importance in the last years. It closely connects physics,
chemistry, material science, biology, and several branches of engineering 
technology. According to its key role in modern science a considerable 
amount of research has been performed
to describe colloidal suspensions from a theoretical point of view and by 
simulations~\cite{Ladd94,Ladd94b,Brady96,Silbert97a,Soga99,Harnau04} as 
well as to understand the particle-particle 
interactions~\cite{DLVO,DLVO2,Alexander84,Grimson91,vanRoij97,Dobnikar03}, 
the phase behavior~\cite{Trappe01,Levin03,Hynninen04,Sator04}, the relevant
processes on the microscale and their influence on macroscopic 
parameters~\cite{Dobnikar03b,Palberg95,Yamamoto06}. Colloidal suspensions
are in fact complicated systems, since different time and length scales
are involved. The particle sizes are on a mesoscopic length scale, i.e.,  in the range 
of nanometers up to micrometers. In systems of particles sized on this length scale
Brownian motion often cannot be neglected. Depending on the particle sizes, 
materials, and concentrations, different interactions are of relevance and often
several of them are in a subtle interplay: electrostatic repulsion,
depletion forces, van der Waals attraction, hydrodynamic interaction,
and Brownian motion are the most important influences. 
The properties of the suspension are strongly depending on the
balance of the microscopic forces between the particles. Especially 
for industrial processes, where one needs to optimize certain material 
properties a detailed understanding of the relevant influences is needed. 
The stability of different microstructures and especially the
clustering process are key properties which are of interest.

In our work we investigate these properties, focusing on \Alumina{} particles 
of diameter $0.37\,\mu$m suspended in water. This is a widely used material in
ceramics~\cite{Oberacker01,Reinshagen06}. We have developed a simulation
code for a Brownian suspension~\cite{Hecht05} and have adjusted the simulation
parameters so that the simulation corresponds quantitatively to a real 
suspension  such that experimental data can be compared directly. The diffusion
coefficient, sedimentation velocity~\cite{Hecht05}, and the viscosity of
the suspension can be reproduced~\cite{Hecht06}. We also have tested the 
influence of polydispersity and found that its influence on the results 
is small. It is much more important to choose the correct mean size of the 
particles\cite{Hecht06}. For \Alumina{} suspensions attractive 
van der Waals forces are important for the behavior of this material. Electrostatic 
repulsion of the charged particles counteracts the attraction and can prevent 
clustering depending on the particle surface charge.
In~\Ref{Hecht06} we have presented how one can relate parameters of
DLVO potentials~\cite{DLVO,DLVO2} with experimental conditions. In the
experiment one can control the \pHvalue{} and the salt concentration.
The latter can be expressed by the ionic strength $I$, which is an effective
concentration of all ions present in the solution. Both, the \pHvalue{}
and the ionic strength, influence the charge of the colloidal particles. 
We have shown that for not too strongly attractive forces one can obtain 
reasonable quantitative agreement with experimental results:
by adjusting the lubrication force, which represents the short range hydrodynamics,
we are able to reproduce rheological data of a real suspension~\cite{Hecht06}.

Three regimes can be identified and plotted in a stability diagram~\cite{Hecht06}, 
which we want to investigate here in more detail: A clustered regime, in which 
particles aggregate to clusters, a fluid-like and stable stable suspension
and a repulsive region, for which the microstructure is similar to the ones 
known from glassy systems.
From our previous work we know that our model works well, even quantitatively, 
in the suspended regime of the stability diagram and close to the borders between 
the different microstructures. In this paper we extend our investigations to 
different \pHvalue{}s, deeper in the clustered regime, and to the repulsive structure. 
We expect to gain insight to the microscopic structure on a qualitative level.

On these grounds we have explored the stability diagram of \Alumina{} 
suspensions. The particles are uncharged close to the so called ``isoelectric
point'' at $\pH = 8.7$. There, for all ionic strengths the particles form 
clusters. For lower \pHvalue{}s particles can be stabilized in solution, 
because they are charged. For low \pHvalue{}s, low salt concentrations,
and high volume fractions a repulsive structure can be found.

In the following section we shortly describe our simulation method.
After that we discuss the criteria we apply to characterize the 
microstructures. We utilize the pair correlation function and the structure factor
to characterize the clustering behavior.
Both of them in principle contain the same information, but we concentrate
on certain peaks in either of them. Each peak in the correlation function and 
in the structure factor corresponds to a certain length scale and we chose
either the correlation function or the structure factor, depending on which of 
the two quantities is more suitable under numerical criterions to observe on a 
given length scale. In the section thereafter we describe our simulation setup. 
In the results section we start with the discussion of the correlation 
function and the structure factor.
Additionally, we evaluate the so-called demixing parameter $\Psi$~\cite{Cates03}.
To characterize the repulsive region we evaluate the mean squared displacement (MSD),
which shows a plateau, if the particle motion consists of different processes
acting on well separated time scales. Finally, the results are summarized in a
stability diagram for our \Alumina{}-suspension. It shows the behavior of the
suspension in an intuitive way and helps to design industrial processes using
this material.

% ==========================================================================================
\section{Simulation method}
% ------------------------------------------------------------------------------------------
\noindent
Our simulation method consists of two parts: a Molecular Dynamics (MD) code, which 
treats the colloidal particles, and a Stochastic Rotation Dynamics (SRD) simulation
for the fluid solvent.

In the MD part of our simulation the colloidal particles are represented 
by monodisperse spheres. We include effective electrostatic interactions and
van der Waals attraction, known as DLVO potentials~\cite{DLVO,DLVO2}, a lubrication 
force and Hertzian contact forces. DLVO potentials are composed of two terms, 
the first one being an exponentially screened Coulomb potential due to
the surface charge of the suspended particles
\be
 \ba{rcl}
  V_{\mathrm{Coul}} &=&
  \pi \epsilon_r \epsilon_0
  \left[ \frac{2+\kappa d}{1+\kappa d}\cdot\frac{4 k_{\mathrm{B}} T}{z e}  
         \tanh\left( \frac{z e \zeta}{4 k_{\mathrm{B}} T} \right)
  \right]^2 \\ \phantom{\pi \epsilon_r \epsilon_0} &&
   \times \frac{d^2}{r} \exp( - \kappa [r - d]),
 \ea
 \label{eq_VCoul}
\ee
where $d$ denotes the particle diameter and $r$ is the distance between the 
particle centers. $e$ is the elementary charge, $T$ the temperature, 
$k_{\mathrm{B}}$ the Boltzmann constant, and $z$ is the valency of the ions of 
added salt. 
$\epsilon_0$ is the permittivity of the vacuum, $\epsilon_r=81$ the relative dielectric
constant of the solvent. $\kappa$ is the inverse Debye length defined by 
$\kappa^2 = 8\pi\ell_BI$, with the ionic strength $I$ and the Bjerrum length 
$\ell_B = 7\angst$. 
The first fraction
in \eq{eq_VCoul} is a correction to the DLVO potential (in the form used 
in~\Ref{Huetter99}), which takes the surface curvature into account and is valid 
for spherical particles~\cite{Trizac02}.

The effective surface potential $\zeta$ is the electrostatic potential 
at the border between the diffuse layer and the compact layer. 
Smoluchowski related it to the electrokinetic 
mobility of the particle as $\mu = \zeta \epsilon_0\epsilon_r / \eta$~\cite{Smoluchowski03}.
The $\zeta$ potential can be related to the \pHvalue{} of the solvent with a so-called
$2pK$ charge regulation model~\cite{Hecht06}.

The Coulomb term competes with the second part of the DLVO potential which is given by the
attractive van der Waals interaction
\be
  \ba{rcl}
  V_{\mathrm{VdW}} $=$ - \frac{A_{\mathrm{H}}}{12} 
     \left[ \frac{d^2}{r^2 - d^2} + \frac{d^2}{r^2} \right. \\ $$
             \;\left. + 2 \ln\left(\frac{r^2 - d^2}{r^2}\right) \right].
  \ea
  \label{eq_VdW}
\ee
$A_{\mathrm{H}}=4.76\cdot 10^{-20}\,\mathrm{J}$ is the Hamaker constant~\cite{Huetter99}. 
The DLVO potentials exhibit two minima, one primary minimum, where the particles touch 
each other. \eq{eq_VdW} diverges where in reality the primary minimum of the potential 
is located. Therefore we model the primary minimum by cutting off the DLVO potentials for 
small separations and substituting them by a parabola which is connected in continuously 
differentiable manner to the DLVO potential. This cutoff is made at distances of several
nanometers, where the potential has alredy reached negative values, i.e. where the 
potential decends to the primary minimum. The primary minimum is separated by a 
potential barrier from the secondary minimum, which occurs at larger particle distances and which is less deep.
For low salt concentrations, i.e., large Debye screening lengths, the electrostatic 
repulsion overcompensates the van der Waals attraction and the secondary minimum disappears.

Long range hydrodynamic interactions are taken into account in the simulation for
the fluid as described below. This can only reproduce interactions correctly down
to a certain length scale. On shorter distances, a lubrication force 
has to be introduced explicitly in the Molecular Dynamics simulation:
\be
  \label{eq_FLub}
  \mathbf{F}_{\mathrm{lub}} = -\vec{v}_{\mathrm{rel},\bot}
    \frac{6 \pi \eta }{r - d}\left(\frac{d}{4}\right)^2, 
\ee
with the relative velocity $\vec{v}_{\mathrm{rel},\bot}$ projected on the connecting 
line of the particle centers. $\eta$ is the dynamic viscosity of the fluid. Our 
implementation of the interaction contains cutoff radii to connect it to the long-range
hydrodynamics and to avoid numerical instability for particles touching each other~\cite{Hecht05, Hecht06}.
\\
To avoid that the particles penetrate each other, we are using a Hertz force described by the potential
\be
  V_{\mathrm{Hertz}} = K (d-r)^{5/2}  \quad  \mathrm{if}  \quad r<d,
\ee
where $K$ is an interaction constant and additionally a damping term
\be
  \mathbf{F}_{\mathrm{Damp}} =  -\vec{v}_{\mathrm{rel},\bot}  \beta  \sqrt{d - r},
\ee
with a damping constant $\beta$.
For the integration of the translational motion we utilize a velocity Verlet algorithm~\cite{Allen87}.

For the simulation of a fluid solvent, many different simulation methods have been proposed:
Stokesian Dynamics (SD)~\cite{Brady88,Brady93,Brady96}, 
Accelerated Stokesian Dynamics (ASD)~\cite{Brady01,Brady04}, 
pair drag simulations~\cite{Silbert97a}, Brownian Dynamics (BD)~\cite{Huetter99,Huetter00}, 
Lattice Boltzmann (LB)~\cite{Ladd94,Ladd94b,Ladd01,Harting04}, 
and Stochastic Rotation Dynamics (SRD)~\cite{Inoue02,Padding04,Hecht05}.
These mesoscopic fluid simulation methods have in common that they make certain approximations
to reduce the computational effort. Some of them include thermal noise intrinsically, or it 
can be included consistently. They scale differently with the number of embedded particles
and the complexity of the algorithm differs largely.

We apply the Stochastic Rotation Dynamics method (SRD) introduced by Malevanets and Kapral~\cite{Malev99,Malev00}. It intrinsically contains fluctuations, is easy to implement, and
has been shown to be well suitable for simulations of colloidal and polymer suspensions~\cite{Inoue02,Padding04,Gompper04,Gompper04b,yeomans-2004-ali,Hecht05,Hecht06}
and very recently for star-polymers in shear flow~\cite{Gompper06}. 
The method is also known as ``Real-coded Lattice Gas''~\cite{Inoue02} or 
as ``multi-particle-collision dynamics'' (MPCD)~\cite{Gompper05}.
It is based on so-called fluid particles with continuous positions and velocities. A streaming
step and an interaction step are performed alternately. In the streaming step, each particle $i$ 
is moved according to
\be
\label{eq_move}
\vec{r}_i(t+\tau)=\vec{r}_i(t)+\tau\;\vec{v}_i(t),
\ee
where $\vec{r}_i(t)$ denotes the position of the particle $i$ at time $t$ and $\tau$ is the 
time step.
In the interaction step the fluid particles are sorted into cubic cells of a regular 
lattice and only the particles within the same cell interact among each other
according to an artificial interaction. The interaction step is designed to exchange
momentum among the particles, but at the same time to conserve total energy and total 
momentum within each cell, and to be very simple, i.e., 
computationally cheap: each cell $j$ is treated independently. First, the mean velocity 
$\vec{u}_j(t')=\frac{1}{N_j(t')}\sum^{N_j(t')}_{i=1} \vec{v}_i(t)$ is calculated.
$N_j(t')$ is the number of fluid particles contained in cell $j$ at time $t'=t+\tau$.
Then, the velocities of each fluid particle in cell $j$ are rotated according to
\be
\label{eq_rotate}
\vec{v}_i(t+\tau) = \vec{u}_j(t')+\vec{\Omega}_j(t') \cdot [\vec{v}_i(t)-\vec{u}_j(t')].
\ee
$\vec{\Omega}_j(t')$ is a rotation matrix, which is independently chosen at random
for each time step and each cell. We use rotations about one of the coordinate axes by 
an angle $\pm\alpha$, with $\alpha$ fixed. 
The coordinate axis as well as the sign of the rotation are chosen at random, 
resulting in 6 possible rotation matrices. To remove anomalies introduced by the 
regular grid, one can either choose a mean free path of the order of the cell size
or shift the whole grid by a random vector once per SRD time step as 
proposed by Ihle and Kroll~\cite{Ihle02a,Ihle02b}.

Three different methods to couple the SRD and the MD simulation have been introduced in the 
literature. Inoue\etal proposed a way to implement no slip boundary conditions on the 
particle surface~\cite{Inoue02}. Padding and Louis very recently came up with full slip
boundaries, where the fluid particles interact via Lennard-Jones potentials with the 
colloidal particles~\cite{Padding06}. Falck\etal\cite{Falck04} have developed a 
``more coarse grained'' method which we use for the simulations of the present paper 
and which we descibe shortly in the following. 

To couple the colloidal particles to the fluid,
the colloidal particles are sorted into the SRD cells and their velocities are included 
in the rotation step. One has to use the mass of each particle---colloidal or fluid particle---as 
a weight factor when calculating the mean velocity 
\bea
\label{eq_rotateMD}
\vec{u}_j(t')&=& \frac{1}{M_j(t')}\sum\limits^{N_j(t')}_{i=1}\vec{v}_i(t) m_i,
\\
\label{eq_rotateMD2}
\mathrm{with}\qquad M_j(t')&=&\sum^{N_j(t')}_{i=1}m_i,
\eea
where we sum over all colloidal and fluid particles in the cell, so that $N_j(t')$ is 
the total number of both particles, fluid plus colloidal ones. $m_k$ is the mass of the
particle with index $i$ and $M_j(t')$ gives the total mass contained 
in cell $j$ at time $t'=t+\tau$. 
In summary, the algorithm for the fluid simulation is described by 
\eqns{eq_move}{eq_rotateMD2}. 
To some of our simulations we apply shear. This is realized by explicitly setting the
mean velocity $\vec{u}_j$ to the shear velocity in the cells close to the border of 
the system. Both, colloidal and fluid particles, are involved in this additional step.
A thermostat is applied to remove the energy introduced to the system by the shear force.
We have described the simulation method in more detail in~\Refs{Hecht05,Hecht06}.

%==========================================================================================
\section{Background}
%------------------------------------------------------------------------------------------
\noindent
\label{sec_backgruond}
In this paper we examine the microstructures obtained in our simulations for different 
conditions. We vary the \pHvalue{} and the ionic strength $I$. The shear rate 
$\dot\gamma$ as an external influence is varied as well. 
We classify the microstructures in three categories: suspended,
clustered, and repulsive. In the suspended case, the particles can move 
freely in the fluid and do not form stable clusters. In the clustered regime
the particles form clusters due to attractive van der Waals forces. These clusters
can be teared apart if shear is applied. In some of our simulations the clusters
are very weakly connected and at small shear rates they are not only broken up 
into smaller pieces, but they dissolve to freely moving individual particles. In this
case, we assign the microstructure to the suspended region, although in complete absence
of the shear flow clusters are formed. At the borders between the different regimes
in fact no sharp transitions can be observed. The DLVO forces rather steadily increase
and compete with the hydrodynamic interactions. Accordingly, in experiments one cannot
observe a sudden solidification, but a steadily increasing viscosity when leaving the 
suspended regime~\cite{Hecht06}. 
\\
Similarly as for attractive forces, repulsive interactions can restrict the mobility
of the particles. If this happens, the mean squared displacement of the particles
shows a pronounced plateau, as it can be found in glassy systems. However, we speak
of a ``repulsive structure'', because the change of the viscosity is not as 
strong as in glasses, where it often changes by many oders of magnitude, when the 
glass transition is approached. In addition, to claim a system shows a glassy behavior
would require to investigate the temperature dependence of a typical time (e.g. particle 
diffusion time) and to show its divergence as the glass temperature is approached.
This is difficult to do in the framework of our simulation model~\cite{Hecht05} and 
therefore we prefer to speak about a ``repulsive structure'' which might
be identified as a colloidal glass in future work. 

In this paper we would like to emphasize the analysis of the microstructure for different
conditions. Our aim is to reproduce a so-called stability diagram by simulations. The 
stability diagram depicts the respective microstructure depending on the \pHvalue{}
and the ionic strength $I$. We apply different numerical tools to analyze the 
microstructure in our simulations and finally arrive at a stability diagram shown 
in \fig{fig_phasediag}, which summarizes the results which we present in the following
sections.

The first tool we apply for that is the pair correlation function 
\be
g(r) = \frac{V}{N^2}\left\langle\sum\limits_i\sum\limits_{j\neq i}\delta(r - r_{ij})\right\rangle,
\ee
(see~\Ref{Allen87} p.55), where $V$ is the volume, $N$ the number of particles and $r_{ij}$
the distance of two particles $i$ and $j$, can be used for a first characterization of the system. It shows maxima at certain typical particle distances, e.g. there is a nearest
neighbor peak, and more complicated structures at larger distances, which we have 
assigned to typical particle configurations for small distances~\cite{Hecht05}.
Here, we use the next-neighbor-peak to analyze our data, which provides information
of the short-range structure of the suspension. To observe the long-range structure
we move on to its complementary quantity: the structure factor, defined by
\be
 S(\vec{k}) = \frac{1}{N}\sum\limits_{l,m=1}^{N} \exp(i \vec{k}\cdot\vec{r}_{lm}),
 \label{eq_Sofk}
\ee
where $N$ is the number of particles, and $\vec{r}_{lm}$ is the vector from particle $l$ to
particle $m$. $i$ denotes the imaginary unit here. The structure factor is defined 
in ${\vec k}$-space and it is related to the pair correlation function in real space by a 
three dimensional Fourier transform:
\be
 S(k) - 1= \int \mathrm{d}\vec{r} \exp(i \vec{k}\cdot\vec{r})\rho g(r),
\ee
with the density $\rho$. Therefore, in principle the structure factor contains the same 
information as the pair correlation function. However, due to numerical reasons and our 
implementation of shear boundary conditions it is easier to observe the long-range 
structure in the structure factor than in the pair correlation function.
For our evaluation it is important that in a finite system \kvector{}s are restricted
to discrete vectors $2\pi\left(\frac{n_x}{L_x},\frac{n_y}{L_y},\frac{n_z}{L_z}\right)$
with $L_{x,y,z}$ being the system size in $x$-, $y$-, and $z$-direction and 
$n_{x,y,z} \in {\mathbb{Z}}$. We evaluate $S(\vec{k})$ for these \kvector{}s and use
$\left|S(\vec{k})\right|$ for further analysis. Since we do not evaluate the anisotropy
of the structure factor in this study, we average over all possible orientations of 
the \kvector{}. We sort the absolute values of the \kvector{}s into intervals and 
average within these intervals over the values of the structure factors. Let us now 
discuss some typical features of the structure factor, corresponding to typical length 
scales present in the system.

One typical length for dense systems is about 
one particle diameter, the distance particle centers have to keep so that they 
do not overlap. The corresponding peak of the structure factor is the nearest 
neighbor peak at $k = \frac{2\pi}{d}$. In a single crystal the peak would be sharp, since the 
particle positions are well defined, whereas in our case of a suspension there is a certain
disorder, which broadens the peak. 

Similar to the nearest neighbor peak another peak can be detected at twice the \kvector{}, 
which corresponds to a distance of one particle radius. This does not necessarily mean that 
there are particles whose centers are only one particle radius apart.
It only means that for a certain \kvector{} the addends in \eq{eq_Sofk} do not cancel out each 
other. 

For small \kvector{}s there is another peak, or in our case an increase of the structure 
factor towards low \kvector{}s. The length scale corresponds to the size of the whole system. 
If large clusters are formed, this can be seen in an increase of the \lowkpeak{},
since the contributions on this length scale do not cancel out each other anymore.

As already mentioned, later in this paper we focus on the \lowkpeak{} which corresponds 
to a length scale of the size of the system. 
%  In principle the information contained in the structure factor 
% is the same as in the pair correlation function, since they are complementary quantities. 
% However, integrating over the \lowkpeak{} numerically yielded better results than integrating
% the pair correlation function for large distances. Therefore we use the nearest neighbor peak 
% of $g(r)$ to observe on the length scale of a particle diameter and we use the \lowkpeak{} 
% of $S(k)$ to see the same on the length scale of the system size.

%==========================================================================================
\section{Simulation Setup}
%------------------------------------------------------------------------------------------
\noindent
In this study the colloidal particles are represented by three dimensional spheres 
of $d = 0.37\,\mu\mathrm{m}$ in diameter. This is the mean diameter of the particles
used in the experiments to which we refer in~\Ref{Hecht06}. We have simulated a small 
volume, $24\,d = 8.88\,\mu\mathrm{m}$ long in $x$-direction, which is the shear direction, 
and $12\,d = 4.44\,\mu\mathrm{m}$ long in $y$- and $z$-direction. We have varied the 
volume fraction between $\Phi = 10\,\%$ (660 particles) and $\Phi = 40\,\%$ (2640 particles). 
Most of the simulations were performed at $\Phi = 35\,\%$ (2310 particles).
We use periodic boundaries in $x$- and $y$-direction and closed boundaries in 
$z$-direction~\cite{Hecht06}. Shear is applied in $x$-direction by moving small zones of particles 
and fluid close to the wall with a given shear velocity. The $xy$-plane is our shear 
plane. For simulations
without shear, to achieve the best comparability, we use the same boundary conditions
and just set the shear rate to $\dot\gamma = 0$. In addition we have performed simulations 
with two different shear rates: with $\dot\gamma = 100/$s and with $\dot\gamma = 500/$s. 
For better comparability to other works the Pecl\'et number
\begin{equation}
  Pe = 6 \pi \eta R^3 \dot\gamma / k_{\mathrm{B}} T,
  \label{eq_def_PecletShear}
\end{equation}
which expresses the importance of Brownian motion with respect to the shear flow, 
is useful. The Pecl\'et numbers in our shear simulations are $Pe = 3$ and $Pe = 15$, 
respectively.

%==========================================================================================
\section{Results and Discussion}
\noindent
First, we focus on simulations without shear, where one can predict intuitively, what 
should happen. Qualitatively the results are similar to our earlier work~\cite{Hecht05}, 
but the quantitative relation between the \pHvalue{} and the potentials is new.
The relation was presented in~\Ref{Hecht06}, but here we apply it to different 
cases and we focus more on the characterization of the microstructure. However, 
given the particle particle interaction potentials, the microstructure in equilibrium 
can be predicted easily, at least on a qualitative level. But, the matter changes and 
gets more sophisticated, when shear is applied and an interplay between shear flow
and particle particle interactions becomes responsible for the resulting microstructure.

\subsection{Correlation function}
\begin{figure}
\mbox{\epsfig{file=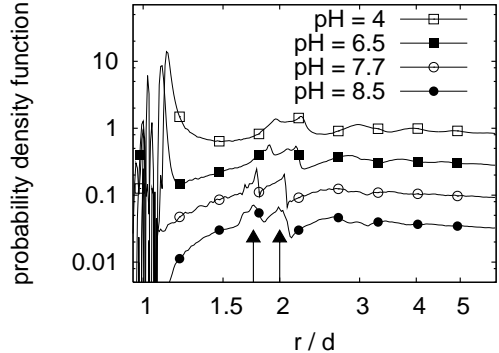,width=\linewidth}}
\caption{Dependence of the particle correlation function on the pH value, $I=3\,$mmol, $\dot\gamma=0/$s $\Phi=35\%$. The plots for four different \pHvalue{}s are shifted against each 
other for better visibility by a factor of 3. For $\pH = 4$ the particles are not clustered. 
Hence the structure at $\frac{r}{d}=2$ is less sharp than in the other three curves of the 
plot and the nearest neighbor peak (at $\frac{r}{d}=1$) is broad.
For $\pH = 6.5$ slight clustering starts, the structures become sharper. For $\pH = 7.7$ 
strong cluster formation is reflected in very sharp structures. For $\pH = 8.5$ electrostatic
repulsion nearly disappears so that no barrier between primary and secondary minimum exists 
anymore. The particles cannot rearrange anymore, and therefore the structures labeled by 
the arrows become smoothened compared to the case of $\pH = 7.7$.}
\label{fig_Corr1}
\end{figure}

\noindent
For constant ionic strength $I = 3\,$mmol/l the local microstructure can be examined 
using the correlation function. Depending on the \pHvalue{} the behavior of the system 
changes from a repulsive structure around $\pH = 4$ to a stable suspension around  
$\pH = 6$ towards a clustered region if the \pHvalue{} is further increased, until the 
isoelectric point is reached at $\pH = 8.7$. There clustering occurs in any case, 
independent on the ionic strength. This can be seen in the structure of the correlation 
function:
electrostatic repulsion prevents clustering (at $\pH = 4$). Particles are suspended, 
and there is no fixed long range ordering in the system. The correlation function 
(\fig{fig_Corr1}) shows a maximum at a typical nearest neighbor distance slightly above 
$\frac{r}{d}=1$ with $d$ denoting the particle diameter, then in the layer of next neighbors 
small correlations can be found 
(at $\frac{r}{d}=2$). For larger distances the correlation function is rather constant. 

When the \pHvalue{} is increased, the surface charge is lower, which at first causes the
particles to approach each other more closely. The maximum of the correlation function
is shifted to smaller distances (see \fig{fig_Corr1}, note that the curves are shifted 
vertically in the plot by a factor of 3 for better visibility.). 
Then, van der Waals attraction becomes 
more important and clustering begins. One can see this in the correlation function 
where a sharp structure at particle distances between 1.5 and 2 particle diameters
occurs. In a solid like cluster the position of the next neighbor is fixed more sharply 
than in the suspension, consequently the nearest neighbor peak becomes sharper, too, and its
height is increased. Close to the isoelectric point ($\pH = 8.7$) the barrier between primary 
and secondary minimum disappears. The particles, once clustered,  cannot rearrange anymore, 
and therefore the correlations to the next neighbors become less sharp again (compare the 
cases of $\pH = 8.7$ and $\pH = 7.7$ in \fig{fig_Corr1} at the positions denoted by the 
arrows).

Instead of varying the \pHvalue{}, one can also vary the ionic strength to achieve 
similar effects. Increasing the ionic strength, experimentally speaking ``adding salt''
decreases the screening length $1/\kappa$ and therefore the attractive forces become
more important: the particles start to form clusters. On the contrary, if the ionic 
strength is decreased--experimentally speaking a dialysis step is performed--the
electrostatic repulsion prevents cluster formation and, for sufficiently strong repulsion, 
long range correlations occur as soon as the range of the repulsion reaches the mean particle distance.

\begin{figure}
\mbox{\epsfig{file=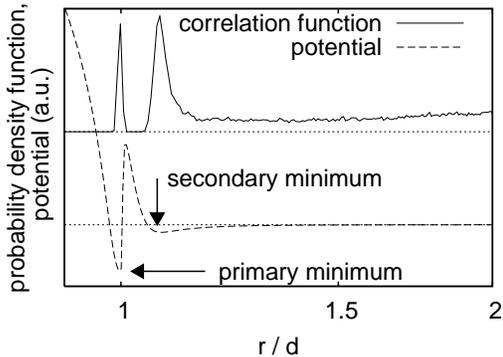,width=\linewidth}}
\caption{Correlation function,  $\dot\gamma=100/$s, $\pH=7$, $I=3\,$mmol/l, $\Phi=35\%$: one 
finds clustering in the primary and the secondary minimum. Note: The vertical axis is
logarithmic, and a constant offset has been added before plotting.
Thereby, the coincidence of the minima in the potential with the maxima in the correlation
function becomes more apparent. The dotted lines denote the position of the base line before 
shifting.}
\label{fig_Corr3}
\end{figure}

In \fig{fig_Corr3} one can see clustering in the primary minimum of the potential as well 
as in the very shallow secondary minimum. The correlation function plotted there is evaluated 
after 1000 SRD time steps in a simulation for $\pH=7$ and $I=3\,$mmol/l at a volume fraction of 
$\Phi=35\%$ sheared with $\dot\gamma=100/$s. Note that the shear flow is not of essential 
importance here, but it supports the particles to overcome the potential barrier between the 
primary and the secondary minimum. The dotted lines in \fig{fig_Corr3} denote the zero line.

A secondary minimum only exists if the screening length of 
the electrostatic repulsion is short enough, i.e., if the ionic strength is large enough.
According to the depth of the potential, clustering in the primary minimum is associated 
with much stronger forces than in the secondary minimum.

\begin{figure}
\epsfig{file=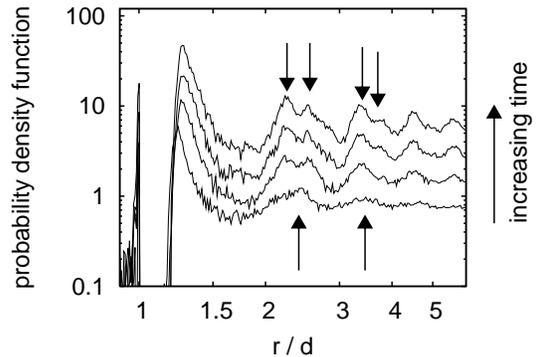,width=\linewidth}%,width=0.9\linewidth}
\caption{Correlation function for $\dot\gamma=500/$s, $\pH=6$, $I=0.3\,$mmol/l, 
$\Phi = 32\,\%$: depending on the simulation time the peaks indicated by the arrows from below
split into two (indicated by the arrows from above), and long range correlations occur 
(for $\frac{r}{d}>4$). 
This reflects the process of layer formation and appearing of a regular (hexagonal) order
in the layers. From bottom to top several states for increasing time are shown.
The plots are shifted vertically by a factor of 2 for better visibility.}
\label{fig_Corr2}
\end{figure}

The effects described up to here can be observed with or without shear qualitatively in an 
analogous manner. If the suspension is sheared clustering occurs at higher
 \pHvalue{}s and the peaks found in the correlation function are slightly broadened,
because the relative particle positions are less fixed. 
But a new feature appears, if a stable suspension of not too high volume fraction is
sheared. Induced by the shear particles arrange themselves in layers. Regular nearest neighbor 
distances in the shear plane cause the correlation function to become more structured even 
for large distances (see \fig{fig_Corr2}). The long range structure of the pair correlation 
function appears after a transient time the particles need to arrange themselves in the 
layered structure. The curves in \fig{fig_Corr2} correspond to the same simulation, but
for increasing time from bottom to top. The arrows from the bottom indicate the position 
of the next-neighbor peak and the next but one neighbors (Since the particles do not touch 
each other, the peaks are located not exactly at $\frac{r}{d}=1, 2, 3, \dots$). Within one 
layer which moves as a whole, a hexagonal particle order appears, which can be seen in the 
occurrence of a peak at $\frac{r}{d} = \sqrt{3}$ times the position of the nearest neighbor 
peak, i.e., the next neighbor peak splits, as indicated by the arrows from above. The same 
applies to the next but one neighbor peak marked by the second set of arrows at $3<r/d<4$.
Shear induced layer formation has been found in both, experiments~\cite{Ackerson88,Ackerson89} and
simulations~\cite{Melrose93,Melrose00,Cohen04}

\begin{figure}
\mbox{\epsfig{file=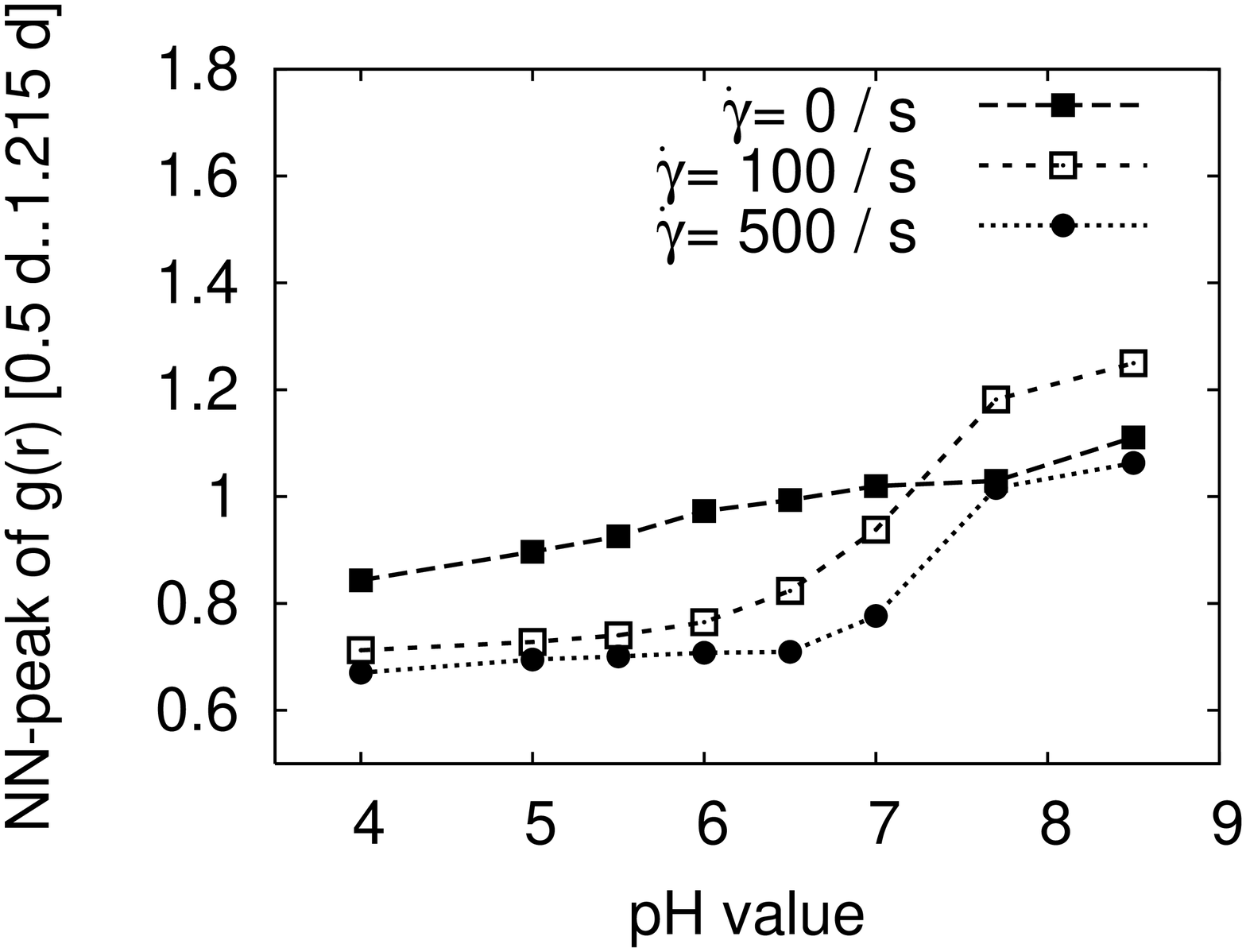,width=\linewidth}}
\caption{Nearest neighbor peak (primary and secondary minimum of the potential) 
of the correlation function $I=3\,$mmol/l, $\Phi = 35\,\%$: 
For low \pHvalue{}s clustering is prevented by the electrostatic repulsion. 
For high \pHvalue{}s the particles form clusters, which is reflected by an increased nearest
neighbor peak. First, shear prevents clustering, then depending on the shear rate, cluster
formation takes place. Low shear rates even support cluster formation at high \pHvalue{}s.
}
\label{fig_CorrNNPeakPh}
\end{figure}

We have integrated over the nearest neighbor peaks, both, the peaks of the primary and the 
secondary minimum, and plotted the integral versus \pHvalue{}
in \fig{fig_CorrNNPeakPh}. We have chosen $I=3\,$mmol/l and $\Phi = 35\,\%$
and three different shear rates: $\dot\gamma = 0, 100$ and $500/$s. We have integrated the 
correlation function for $r < 1.215\,d$, where for all \pHvalue{}s the potential in the 
secondary minimum has a value of $-\frac{1}{2} k_{\mathrm{B}}T$. In other words, we have 
captured the primary and the secondary minimum of the potential for this plot.
For low \pHvalue{}s clustering (in the secondary minimum) is only possible for low shear
rates. For high shear rates, the hydrodynamic forces do not allow the formation of stable 
clusters. For rising \pHvalue{}s the clustering increases, first for the un-sheared suspension,
at higher \pHvalue{}s for low shear rates ($\dot\gamma = 100/$s) and finally for high 
shear rates ($\dot\gamma = 500/$s). Remarkably, for $\pH > 7.5$ the curve for $\dot\gamma = 100/$s
shows stronger cluster formation than the other ones. Particles are brought together by 
the shear flow, so that compared to the case of no shear, the clustering process is 
supported here. On the other hand, the shear stress may not be too strong, because otherwise
the clustering process is limited by the shear flow again (for $\dot\gamma = 500/$s the clustering is less pronounced than for $\dot\gamma = 100/$s).

%------------------------------------------------------------------------------------------
\subsection{Structure Factor}
\noindent
The pair correlation function can be used to characterize the local order of the 
microstructure on the length scale of the particle size. However, to do the characterization
on the length scale of the system size, we use the structure factor $S(\vec{k})$.
In principle we could integrate the correlation function $g(r)$ over an interval for
large $r$. However, due to our implementation of shear (see~\Ref{Hecht06}), we have
to close the boundary in $z$-direction. Therefore we already find restrictions for 
$r$ around half the extension of the system in this direction. We find that it is 
easier to handle these finite size restrictions by moving on to the structure 
factor. There, we always find a \lowkpeak{}, even if no cluster formation takes 
place in the simulation. In contrast to experiments, where this peak only appears
for clustered samples, it reflects the presence of a typical length of the system 
size in the simulation. Namely, the finite size of the simulation volume is the 
typical length, which appears in the structure factor by the \lowkpeak{}. 
This only produces a constant offset when integrating over the \lowkpeak{}
as we are going to do below in this section.
Thus, it is easy to handle the influence of the finite system size in $\vec{k}$-space.

\begin{figure}
\mbox{\epsfig{file=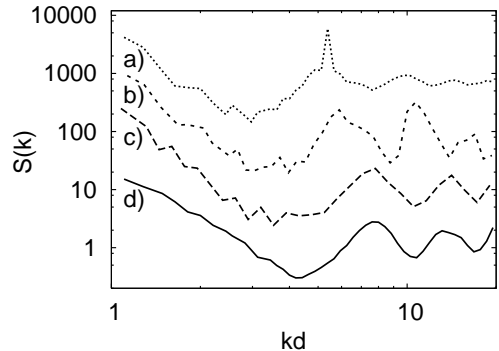,width=\linewidth}}
\caption{Structure factor for some selected examples, with $\pH = 6$ fixed for all plots: 
$\dot\gamma=500/$s, $I = 0.3\,$mmol/l: 
a) $\Phi = 20\%$ and b) $\Phi = 35\%$ , 
$\dot\gamma = 0$, $I = 25\,$mmol/l: 
c) $\Phi = 40\%$ and d) $\Phi = 10\%$ .
The curves are shifted vertically for better visibility. 
In case a) ten layers can be identified in the system, resulting in the strong
peak close to 5. But, since the particles in the layers can still move freely,
there is no \highkpeak{}.
In case b) layers are formed, but particles are moving from one layer to 
the other, disturbing the flow. As a result the nearest neighbor peak is much broader.
Due to the structure in the layers, a  \highkpeak{} appears.
In case c) the interaction is strongly attractive, hence the particles approach 
each other and the nearest neighbor peak is shifted to higher \kvector{}s.
In case d) the volume fraction is much less. The  slope of the \lowkpeak{} is 
much flatter, which  depicts that the cluster is fractal.
}
\label{fig_Sofk}
\end{figure}

\begin{figure}
 a)\epsfig{file=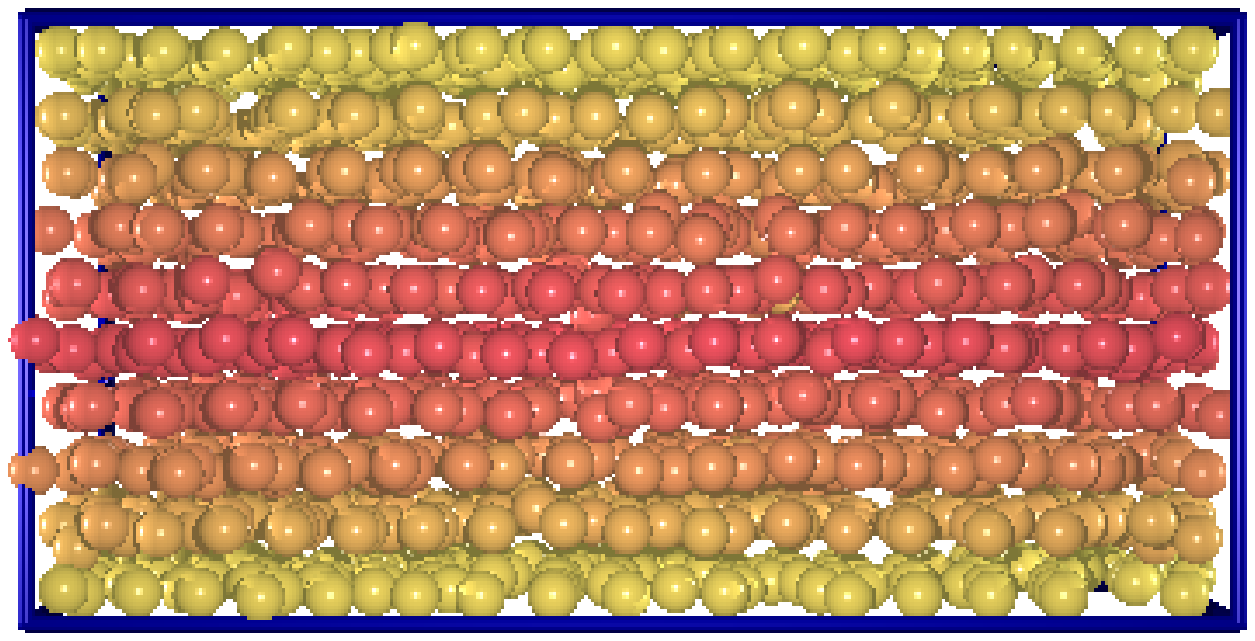,width=0.45\linewidth}
 b)\epsfig{file=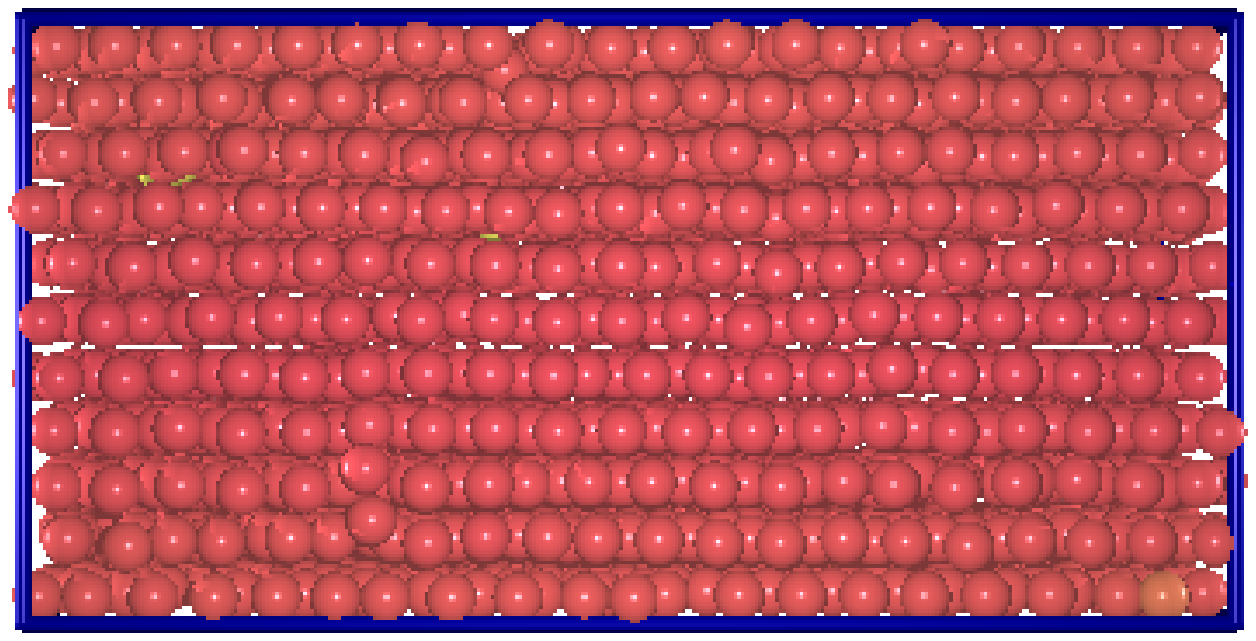,width=0.45\linewidth}
 c)\epsfig{file=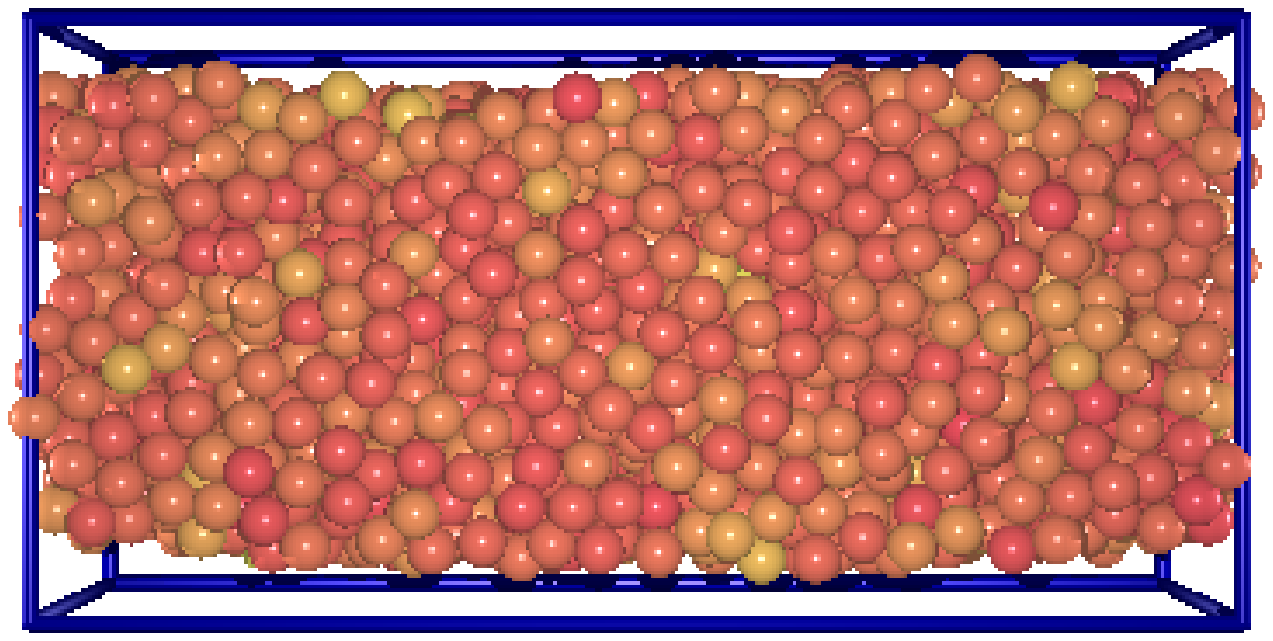,width=0.45\linewidth}
 d)\epsfig{file=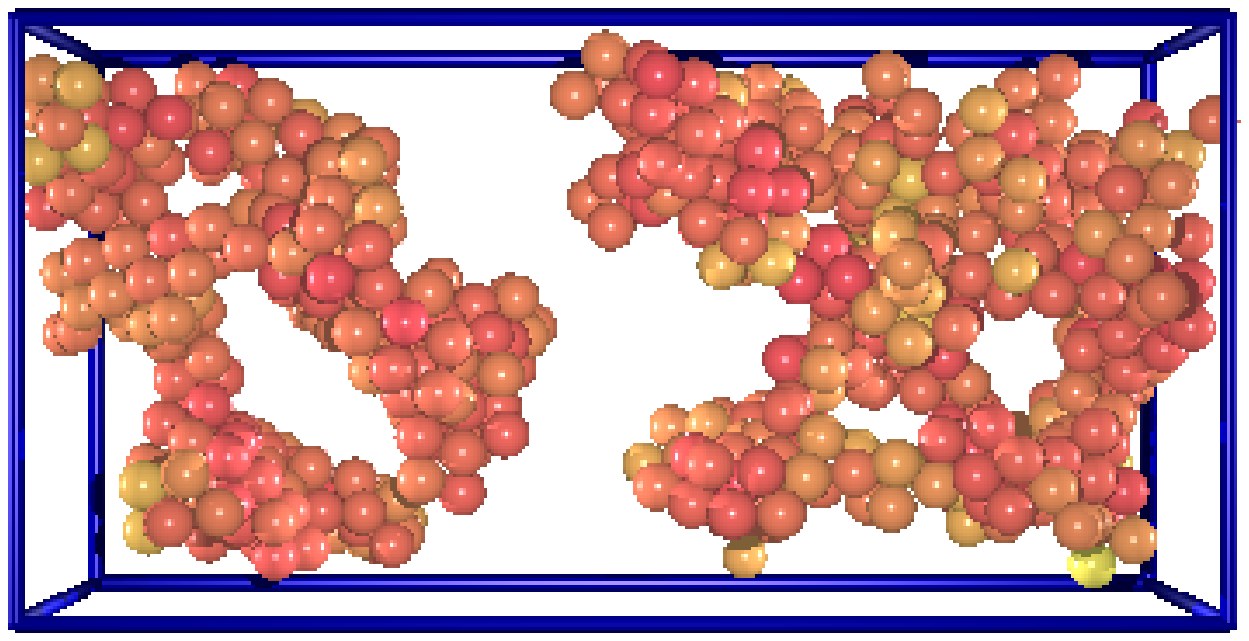,width=0.45\linewidth}
\caption{(Color online) Snapshots of the systems analyzed in \fig{fig_Sofk}:
In case a) one can see the layers resulting in the sharp peak in the structure factor.
In case b) the layers are packed closer due to the higher volume fraction. Collisions 
between particles of neighboring layers happen more frequently.
In case c) one big cluster is formed. The particles are packed densely.
In case d) the fractal nature of the system can be seen directly.
}
\label{fig_snapshots}
\end{figure}

\begin{figure}
\mbox{\epsfig{file=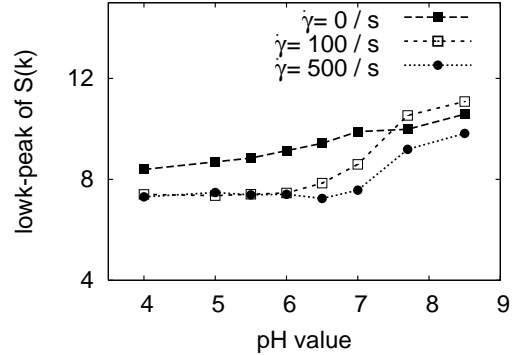,width=\linewidth}}
\caption{\lowkpeak{} for different \pHvalue{}s and different shear rates. The ionic strength $I$ 
is kept constant at $I = 3\,$mmol/l and the volume fraction is always $\Phi = 35\%$. 
For $\dot\gamma=0/$s the particles tend to cluster in the
secondary minimum of the potential. This clustering can easily be broken up, if shear is applied.
If the \pHvalue{} is increased, shear cannot prevent cluster formation anymore. At low 
shear rates ($\dot\gamma = 100/$s) clustering is even enhanced, since the particles are 
brought closer to each other by the shear flow. Note that the offset of the plots 
reflects the finite size of our system combined with the closed boundary conditions 
(see the text).}
\label{fig_lowk}
\end{figure}

In \fig{fig_Sofk} we have plotted several typical structure factors of our simulations. 
For these plots the \pHvalue{} is fixed to $\pH = 6$. The cases a) and b)
are sheared with $\dot\gamma = 500/$s at an ionic strength of $I = 0.3\,$mmol/l. 
In case a) the volume fraction $\Phi = 20\%$ is relatively low. Therefore the particles
can arrange themselves in layers parallel to the shear plane, which move relatively
independently in the shear flow. They have a certain distance fixed in space and time.
This can be seen in a sharp peak at a dimensionless \kvector{} of $k = 5.2$, which 
corresponds to a distance of 1.2 particle diameters. In fact, this is exactly the
distance between two neighboring layers, as one can easily verify by counting the layers
in a snapshot of the system (\fig{fig_snapshots}a)). The particles in the layers 
do not have a fixed distance and therefore no \highkpeak{} can be observed.
\\
For case b) the volume fraction is increased to $\Phi = 35\%$. The particle layers
are packed more densely and therefore the interactions between one layer and the 
neighboring one become relevant. Particles jump from one layer to the other,
which disturbs the flow and therefore the distance between the layers is not 
fixed anymore. The sharp peak on top of the nearest neighbor peak disappears. 
Instead of that, in each layer a regular hexagonal order appears and therefore
the \highkpeak{} is much more pronounced.
\\
In case c) the ionic strength is increased to $I = 25\,$mmol/l. The inter-particle 
potentials are attractive enough that aggregation takes place. In this simulation
we did not apply shear, therefore one finds only one big cluster in the system 
(compare \fig{fig_snapshots}c)). In the cluster the particles are packed more densely and
consistently the nearest neighbor peak in the structure factor is shifted to larger
\kvector{}s. The volume fraction is $\Phi = 40\%$ in this case. 
\\
In case d) the volume fraction is decreased to $\Phi = 10\%$. The particles still 
form clusters, but their mobility is not high enough to create one compact cluster.
The system has a fractal structure (see \fig{fig_snapshots}d)). This can be seen in the 
structure factor as well: The slope of the \lowkpeak{} is flatter in this case compared
to cases a)--c). A flatter slope of the \lowkpeak{} is typical for structure factors
of fractal objects. The fractal dimension of the cluster extracted from the slope 
of the \lowkpeak{} is $2.5$. In experiments this relation is often used to determine the 
fractal dimension of a sample: Lattuada et~al.~\cite{Lattuada01} have evaluated the fractal 
dimension of agglomerates of latex particles from the slope of the structure factor.
McCarthy et~al.~\cite{McCarthy98} give an introduction to scattering 
intensities at fractal objects, without mentioning the structure factor, but their 
arguments refer to the contribution of the structure factor on the scattering intensity.
The underlying mechanism which is responsible for these structures is cluster cluster 
aggregation\cite{Soga98}.
\\
In \fig{fig_lowk} we show the dependence of the \lowkpeak{} of the structure factor
on the \pHvalue{}. Here we have integrated over dimensionless \kvector{}s smaller than 3
which means, we have captured structures larger than twice a particle diameter. A large
integral over the \lowkpeak{} is due to a large inhomogeneity in the system. In 
one part of the system particles are present and in the other part not. In other words, we 
observe the process of cluster formation on a length scale of the system size. Without
shear, particles cluster in the secondary minimum for all \pHvalue{}s. If the system 
is slightly sheared ($\dot\gamma = 100/$s) clustering is suppressed for low \pHvalue{}s. 
Starting at $\pH = 6$ cluster formation starts and is even supported by the shear 
flow for \pHvalue{}s larger than 7.5. For large shear rates ($\dot\gamma = 500/$s) 
cluster formation is suppressed by the shear flow. By analyzing the \lowkpeak{} of the structure 
factor one observes on the length scale of the system size. The same behavior of the system 
can be seen by analyzing the pair correlation function, as we have already shown in \fig{fig_CorrNNPeakPh}.
In that case one analyzes the number of nearest neighbors, that means, one observes 
the length scale of a particle diameter. Nevertheless, both graphs show the same behavior 
of the system, i.e., we have a consistent picture of the cluster formation process
on the length scale of the nearest neighbors and on the length scale of the system size.

Thus we have confirmed that the cluster formation process is not limited to length scales 
smaller than our system size. This is reflected
especially by the transition between $\pH = 7 - 8$ and its shear rate dependence in the 
plots in \fig{fig_lowk} and \fig{fig_CorrNNPeakPh}. There is a strong
similarity of the two plots, which are obtained by two evaluation methods referring to 
two different length scales. This confirms that the 
plots do not only reflect how clusters are formed on the respective length scale, but
that the clustering process is a phenomenon which can be observed on \emph{any} length
scale by applying a suitable method to characterize it.

%------------------------------------------------------------------------------------------
\subsection{Density Inhomogeneity}
\label{sect_demixing}

Another way to observe the cluster formation process is provided by the
``demixing parameter'' defined in \Ref{Cates03} as follows: the system is divided into $n^3$ cubes and
the particle density $\rho_k$ is evaluated in each cube. Then the demixing parameter 
is the mean squared deviation of the density
\be
 \Psi_n = \sum\limits_{k=0}^{n^3} (\rho_k-\bar\rho)^2,
\ee  
where $\bar\rho$ is the mean density. For our elongated system we modify this 
definition and use $2n$ cubes in $x$-direction, resulting in $2n^3$ cubes in total.
\begin{figure*}
\epsfig{file=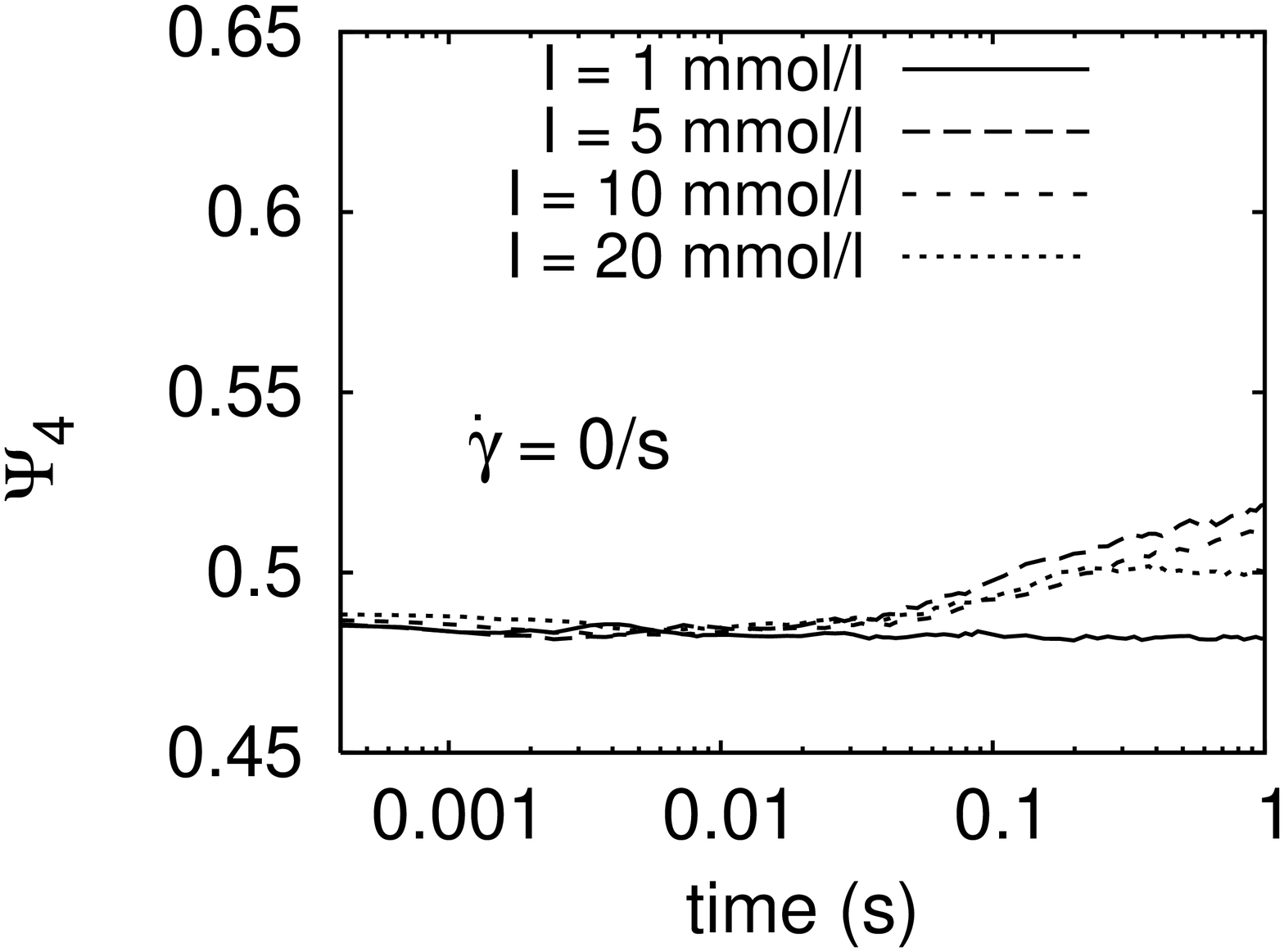,width=0.3\linewidth}
\epsfig{file=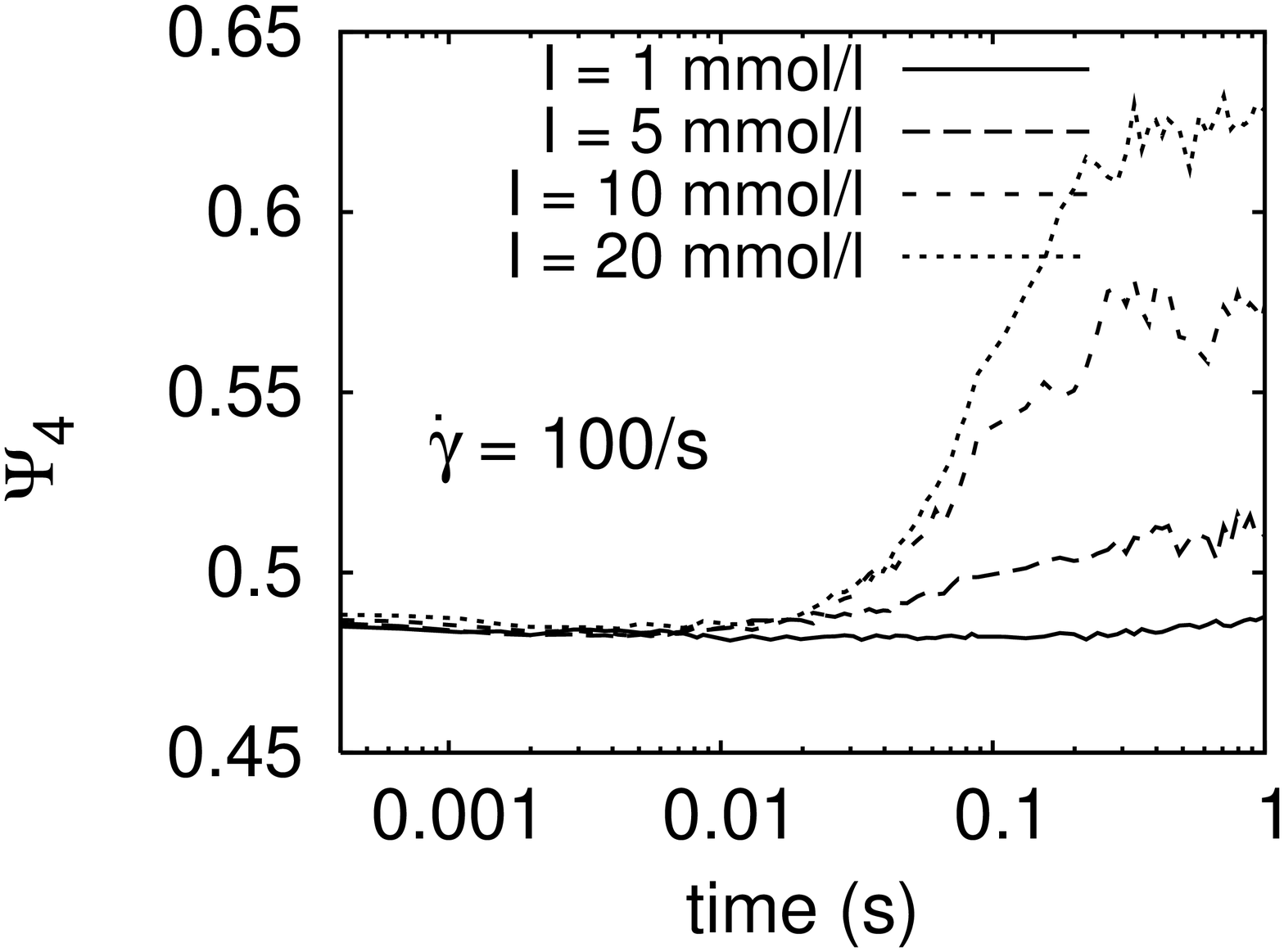,width=0.3\linewidth}
\epsfig{file=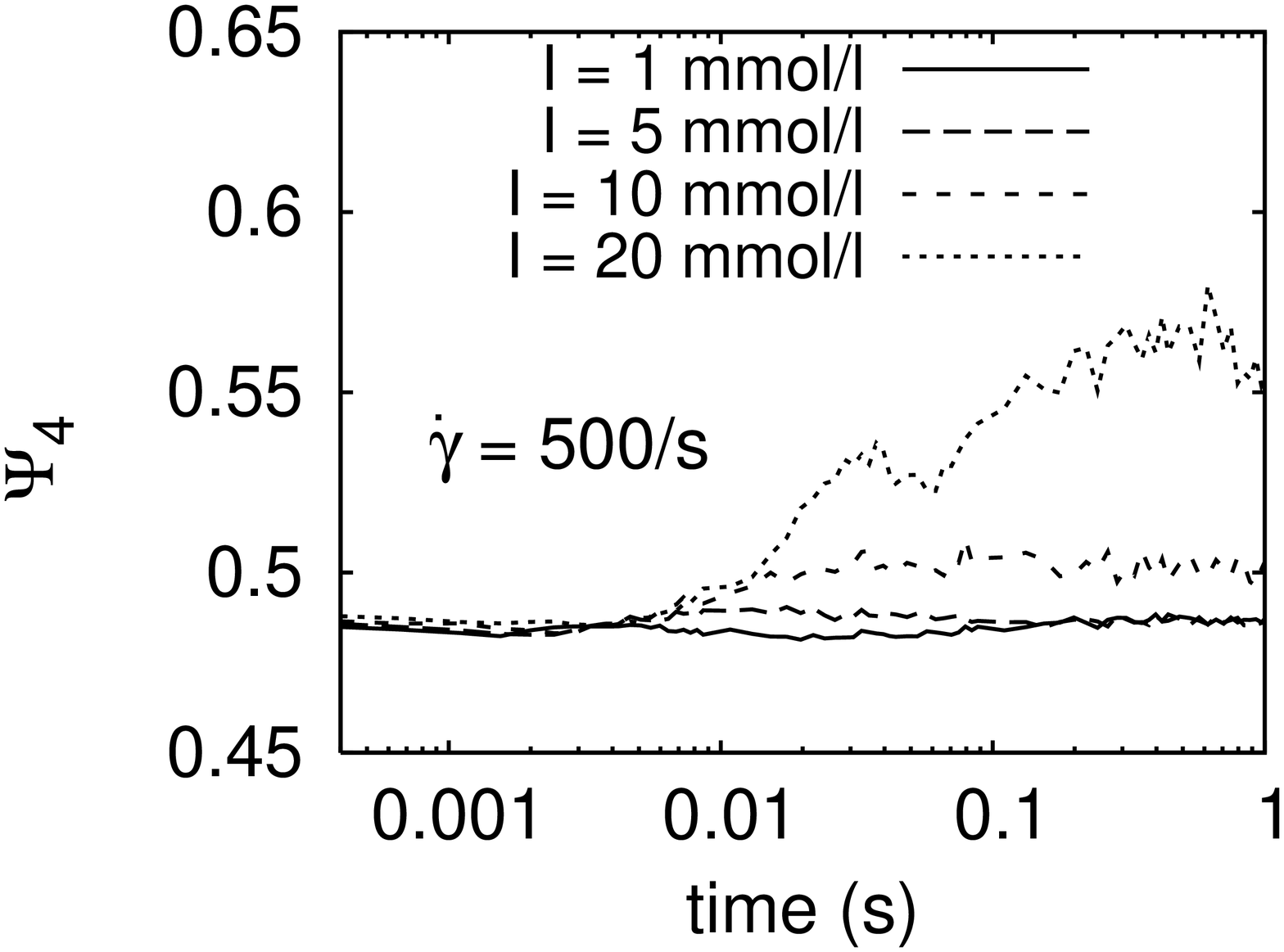,width=0.3\linewidth}
\caption{Demixing parameter for constant $\pH = 6$ and volume fraction $\Phi = 35\,\%$ 
for three different shear rates: $\dot\gamma = 0/$s (left), 100/s (center) and 500/s (right).
Without shear only very weak demixing takes place. For $\dot\gamma = 100\invsec$ the strongest
demixing can be observed for the system with the highest ionic strength $I = 20\mmoll$. 
For lower ionic strengths the effect decreases until it disappears completely at $I = 1\mmoll$,
where the repulsive regime is reached. If the shear rate is further increased 
($\dot\gamma = 500\invsec$), the clusters are less stable so that the system becomes less 
inhomogeneous. For $I = 5\mmoll$ the demixing is even suppressed completely.
}
\label{fig_demixing}
\end{figure*}
The demixing parameter %$\Psi$ shown in \fig{fig_demixing} 
implicitly contains information about the density of the 
clusters. It is the density fluctuation of the whole system. If the clusters 
are more compact, voids have to appear and the demixing parameter increases,
since the distribution of the local density is broadened thereby.

In \fig{fig_demixing} we have plotted the demixing parameter $\Psi_4$ versus time for three 
different shear rates and four different ionic strengths in each plot. The \pHvalue{} is 
kept constant at $\pH = 6$ and the volume fraction at $\Phi = 35\,\%$. Without shear $\Psi_4$ is nearly constant. 
Even if a cluster is 
formed, it does not move, so that the average density in the boxes does not
change much. For lower volume fractions the time dependence without shear is stronger. 
If shear is applied the clusters are deformed so that stronger inhomogeneities appear. 
Depending on the ionic strength the demixing parameter increases with cluster formation 
(high ionic strength) or still stays constant when the interactions are repulsive 
(low ionic strength). For $\dot\gamma = 100\invsec$ the strongest cluster formation is 
achieved for the highest ionic strength $I = 20\mmoll$. If the ionic strength is decreased,
the clustering effect decreases as well, until it disappears completely at $I = 1\mmoll$, 
where the repulsive regime is reached. To see the shear rate dependence, compare the 
plots for  $\dot\gamma = 100\invsec$ and $500\invsec$. If the shear rate is increased,
the cluster formation starts faster because the shear flow brings the particles faster in 
contact, but the resulting inhomogeneity of the final state is less than for lower shear 
rates. This 
%shows again the crossover for nonzero shear rates in \fig{fig_CorrNNPeakPh} and \fig{fig_lowk}. 
effect can be seen in \fig{fig_CorrNNPeakPh} and \fig{fig_lowk}, too. There, also the intensity of 
the respective peaks for $\dot\gamma=500\invsec$ is less than for  
$\dot\gamma=100\invsec$. This shows again that large shear rates can inhibit cluster formation, whereas moderate shear rates can support the clustering process (the plots in 
\fig{fig_demixing}\,b) for $\dot\gamma=100\invsec$ are steeper than the ones in 
\fig{fig_demixing}\,a) for $\dot\gamma=0\invsec$.)
In addition to the information already contained in other quantities we have presented, 
\fig{fig_demixing} shows the time evolution. One can see three regimes of time evolution:
for very short times all plots are nearly constant. This means that the cluster formation
has not yet started. Then, the plots for attractive potentials raise, which shows the onset
of cluster formation. For large times the slope of the plots decreases, which reflects that
the clusters have been formed and no more ``demixing'' takes place. Without shear 
(see \fig{fig_demixing}\,a)\,), it is remarkable that for the largest ionic strength the 
growth of the demixing parameter first becomes constant after 0.3\,s, whereas for the other
cases where clustering occurs, it grows further. This reflects that in the case of $I = 20\mmoll$ 
the attraction is that strong that no reordering of the particles is possible anymore.

\begin{figure}
\mbox{\epsfig{file=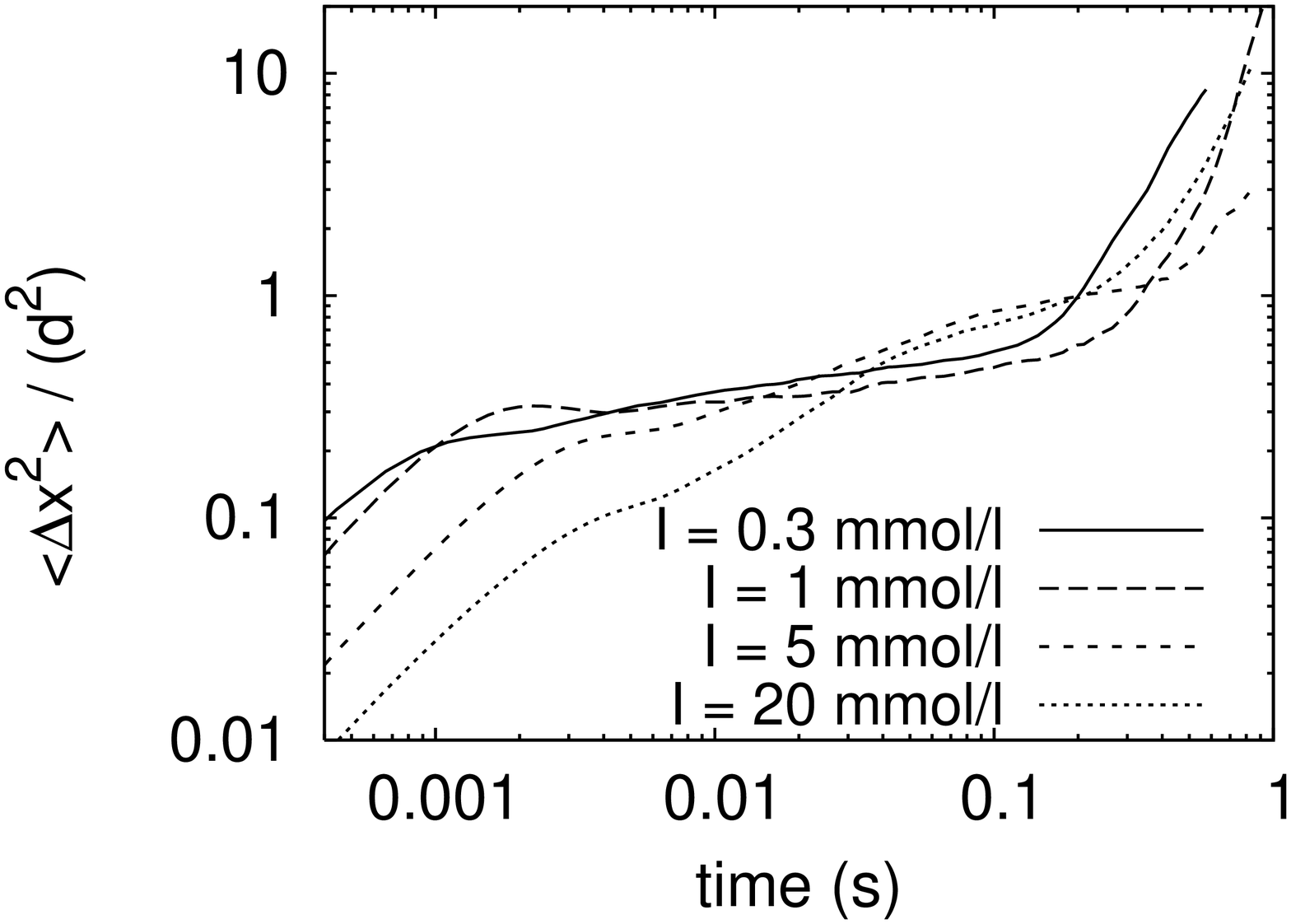,width=\linewidth}}
\caption{Mean squared displacement at $\pH = 6$ for different ionic strengths, without shear.
One can see a ballistic regime for short times, a central plateau and a collective long time
movement which can be a movement of a whole cluster or cage escape events of single particles.
Depending on the ionic strength, the central plateau is more or less pronounced. A comparison 
of the plateau for different simulations can be used to decide, if a certain state belongs to the repulsive region of the stability diagram. A state well in the liquid microstructure should be 
used as a reference for the comparison. }

\label{fig_MSQ}
\end{figure}

%------------------------------------------------------------------------------------------
\subsection{Repulsive Regime}
\label{sec_glassy}
To characterize the repulsive regime, we evaluate the mean squared displacement
for the particles. In \fig{fig_MSQ} we plot the mean squared displacement for 
different ionic strengths. The \pHvalue{} is kept constant at $\pH = 6$ and the volume
fraction is $\Phi = 35\,\%$ for this plot. Three different regimes can be identified.
For very short times, the ballistic regime: particles move on short distances without 
a notable influence by their neighbors. The distances are in the order of some percent
of the particle diameter and the times are a few SRD time steps. For larger times the 
particles interact with their neighbors and therefore their mobility is limited due to 
collisions with the neighbors. This is reflected in the mean squared displacement by a 
plateau of reduced slope, which is the more pronounced the more the mobility of the 
particles is restricted. For even larger 
time scales collective motion starts, i.e., clusters or groups of particles move, or single 
particles can escape from a cage formed by its neighbors. Depending on the ionic strength
different effects are important and thus the shape of the curve is different. 
For large ionic strengths the particles form clusters 
and these clusters may drift or rotate in the system. Then the collective motion is 
more dominant and the mean squared displacement grows faster than in single particle 
diffusion. The mean squared displacement does not show a plateau, then. But in the repulsive 
regime, the neighbors limit the motion of the particles, and the slope of the plateau
is flatter, i.e., the plateau is even more pronounced, compared to the suspended case. 
In the repulsive regime the particles tend to arrange themselves in layers when shear
is applied~\cite{Hecht06} and long range correlations can be found in un-sheared 
systems~\cite{Hecht05}.

%------------------------------------------------------------------------------------------
\subsection{Stability Diagram}
\begin{figure}
\mbox{\epsfig{file=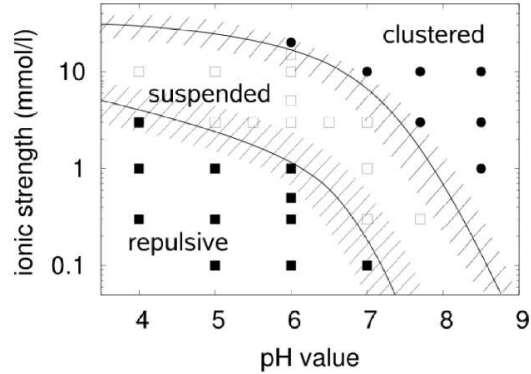,width=\linewidth}}
\caption{Stability diagram (plotted for $\Phi = 35 \%$ and without shear): 
depicting three regions:
a clustered region (filled circles),
a suspended regime (open squares), and
a repulsive structure (filled squares).
In the clustered region particles aggregate which leads to inhomogeneity 
in the system. In the suspended regime, the particles are distributed
homogeneously in the system and they can move freely. In the repulsive 
regime the mobility of the particles is restricted by electrostatic 
repulsion exerted by their neighbors. As a result they arrange in a local 
order which maximizes nearest neighbor distances.
The borders between the regimes are not sharp. They depend on the shear 
rate and on the volume fraction. Therefore we have indicated the crossover 
regions by the shaded patterns. The lines are guides to the eye.}
\label{fig_phasediag}
\end{figure}

\noindent
The results of the investigations presented in this paper can be summarized in a 
stability diagram for our \Alumina{}-suspension (\fig{fig_phasediag}).
Three different microstructures can be identified: a repulsive structure, a suspended 
region and a clustered region. In contrast to our previous 
work~\cite{Hecht05,Hecht06}, we have explored the parameter space more in the 
repulsive regime and deeper in the clustered region. We use the mean squared
displacement, the demixing parameter $\Psi$~\cite{Cates03}, the correlation function, 
and the structure factor, to decide to which of the three microstructures a certain 
point in the stability diagram belongs. However, the borders between the 
regions are not sharp and they depend on the shear rate. We have indicated the crossover 
regions by the shaded patterns in the stability diagram. If the volume fraction 
is decreased, the region of the repulsive structure becomes smaller.

To decide if a state is in the suspended region or in the repulsive one of the 
stability diagram, we have compared the plots of the mean squared displacement
for the simulations without shear. If the plateau was pronounced there, we have 
counted the state among the repulsive regime.  As a second criterion one can 
compare the pair correlation function. If there are long range correlations even 
though the system is not sheared, then the microstructure is the repulsive one. 
Finally, the shear force can be used to localize the border to the repulsive regime.
For a given  shear rate and a fixed volume fraction, the shear force depends on the 
particle interactions. If the shear force increases compared to a state well in the 
suspended regime, the motion of the particles is blocked by the electrostatic
interaction in the repulsive regime.

As we have mentioned in \sect{sec_backgruond}, without shear weak clustering 
can be seen in the suspended case as well, since there is no barrier for the particles 
to enter the secondary minimum of the DLVO potential, but the clusters can be broken 
up again very easily. Thus, to decide, if a state belongs to the clustered or
to the suspended regime, we first study the snapshots of the system. If we 
see no clusters there, the clustered regime can be excluded. But, if we see
clusters, we next check the density of the clusters and the onset time of the
increase of the demixing parameter $\Psi$, which is a measure for the time it 
takes to form clusters. Both, the density and the time are indications for the 
stability of the clusters. If they grow slowly and their density is low, we count 
the state to the suspended regime. The demixing parameter $\Psi$ shown 
in \fig{fig_demixing} implicitly contains information about the density of the 
clusters. It is the density fluctuation of the whole system. If the clusters 
are more compact, voids have to appear and the demixing parameter increases,
since the distribution of the local density is broadened thereby.
The stability diagram obtained by these criteria is consistent with the 
results of the simulations with shear flow, shown in \fig{fig_CorrNNPeakPh} 
and \fig{fig_lowk}. Especially, the increased cluster formation for $I=3\mmoll$
starting between $\pH = 7 - 8$ is reflected in an increased nearest neighbor peak 
in \fig{fig_CorrNNPeakPh}, and \lowkpeak{} in \fig{fig_lowk} respectively, and
in a border between suspended and clustered regime in \fig{fig_phasediag}.
The repulsive structure for $\pH = 4$ and $I = 3 \mmoll$ can not be recognized 
in  \fig{fig_CorrNNPeakPh} and \fig{fig_lowk}, but in a pronounced layer formation.
  
%==========================================================================================
\section{Summary and Outlook}
%------------------------------------------------------------------------------------------
\noindent
We have simulated colloidal particles in shear flow and investigated how the clustering 
process due to attractive DLVO potentials is affected  by the hydrodynamic forces. We 
find a consistent behavior on different length scales. The nearest neighbor peak of 
the pair correlation function has been used to observe the direct neighborhood of the 
particles and the \lowkpeak{} of the structure factor to keep track of the 
length scales up to the system size. In both cases a suppression of the 
cluster formation by the shear flow can be seen at low \pHvalue{}s. For 
large \pHvalue{}s low shear rates even support the clustering process.
In contrast, for high shear rates it suppresses the cluster formation.
We have evaluated the mean squared displacement and the demixing parameter
$\Psi$~\cite{Cates03} in order to draw the stability diagram as given in \fig{fig_phasediag}. 
To our knowledge this stability diagram for \Alumina{} suspensions is reproduced 
quantitatively for the first time from simulations. It helps to predict the behavior of 
a real suspension. Our findings on the cluster formation process suggest
that soft stirring can enhance the cluster formation in industrial processing
of this material. Further investigations can be carried out on the fractal dimension 
and its dependence on the experimental conditions. The \lowkpeak{} of the structure
factor can be used for that. The cluster size distribution could as well deliver 
interesting insights useful to design industrial processes. Apart from that one can
apply our algorithm to different materials. To do so, one has to change the interaction
constants and especially for the electrostatic repulsion the calibration we have 
presented in \Ref{Hecht06} are necessary. We expect that the stability
diagram does not change qualitatively, but the position of the borders will be different.

%\begin{acknowledgement}
%\textbf{Acknowledgments.}
\subsection*{Acknowledgments}
\noindent
This work has been financed by the German Research Foundation (DFG) within the project
DFG-FOR 371 ``Peloide''. We thank G.~Gudehus, G.~Huber, M.~K\"ulzer, L.~Harnau, M.~Bier, 
J.~Reinshagen, and S.~Richter for valuable collaboration. 
\\
This work is resulting in large parts from the collaboration with the group of 
A.~Coniglio, Universit\`{a}  di Napoli ``Federico II'', Naples, Italy. 
M.~Hecht thanks him and his group for his hospitality and 
for the valuable support during his stay there. M.~Hecht thanks the DAAD for the 
scholarship (Doktorandenstipendium) which enabled him the stay.
\\
The computations were performed on the IBM p690 cluster at the Forschungszentrum 
J\"{u}lich, Germany and at HLRS, Stuttgart, Germany. 

%\end{acknowledgement}
%\bibliographystyle{short}
\bibliographystyle{unsrt}

\end{document}